%% file: arxiv.tex
\documentclass[prd,superscriptaddress,aps,amsfonts,showpacs,longbibliography,notitlepage,twocolumn,nofootinbib]{revtex4-1}
\usepackage[dvipsnames]{xcolor}
\usepackage{multirow}
\usepackage{tabularx}
\usepackage{nicematrix}

\include{preamble}

\newcolumntype{Y}{>{\centering\arraybackslash}X}
\newcolumntype{M}[1]{>{\centering\arraybackslash}m{#1}}

\newcommand{\comment}[1]{}

\begin{document}
\title{Entangling four logical qubits beyond break-even in a nonlocal code}

\author{Yifan Hong}
\email{yifan.hong@colorado.edu}
\affiliation{Department of Physics and Center for Theory of Quantum Matter, University of Colorado, Boulder, CO 80309, USA}

\author{Elijah Durso-Sabina}
\affiliation{Quantinuum, Broomfield, CO 80021, USA}

\author{David Hayes}
\affiliation{Quantinuum, Broomfield, CO 80021, USA}

\author{Andrew Lucas}
\affiliation{Department of Physics and Center for Theory of Quantum Matter, University of Colorado, Boulder, CO 80309, USA}

\begin{abstract}
Quantum error correction protects logical quantum information against environmental decoherence by encoding logical qubits into entangled states of physical qubits.  One of the most important near-term challenges in building a scalable quantum computer is to reach the break-even point, where logical quantum circuits on error-corrected qubits achieve higher fidelity than equivalent circuits on uncorrected physical qubits. Using Quantinuum's H2 trapped-ion quantum processor, we encode the GHZ state in four logical qubits with fidelity $ 99.5 \pm 0.15 \% \le F \le 99.7 \pm 0.1\% $ (after postselecting on over 98\% of outcomes). Using the same quantum processor, we can prepare an uncorrected  GHZ state on four physical qubits with fidelity $97.8 \pm 0.2 \% \le F\le 98.7\pm 0.2\%$.  The logical qubits are encoded in a $\llbracket 25,4,3 \rrbracket$ Tanner-transformed long-range-enhanced surface code.  Logical entangling gates are implemented using simple swap operations. Our results are a first step towards realizing fault-tolerant quantum computation with logical qubits encoded in geometrically nonlocal quantum low-density parity check codes.
\end{abstract}

\date{\today}

\maketitle

\emph{Introduction.}--- It is widely believed that fault tolerance will be necessary for quantum computers to outperform classical computers on large-scale problems. Decoherence, if left ignored, will eventually destroy entanglement and any quantum advantage in computation. A critical milestone for the development of a quantum computer will, therefore, be the protection of logical qubits and the implementation of logical gates whose error rates are below that of physical qubits and circuits which are not error-corrected.  When logical qubits are more robust than physical qubits, we have reached \emph{break-even}.

Since the discovery of topological quantum codes \cite{Kitaev_2003, Bombin_2006} and their promise to preserve quantum information for arbitrarily long times with a relatively high tolerance for errors \cite{Dennis_2002, landahl2011}, there have been immense efforts to implement such codes in current architectures \cite{Postler2021,Ryan-Anderson_2021, Google_2023, Bluvstein_2023, silva2024}. Unfortunately, the resource requirements for large-scale, fault-tolerant quantum computation using topological codes with current hardware specifications may be enormous \cite{Gidney_2021}. 

More generally, it is known that these overheads cannot be substantially improved in planar architectures (codes) with only nearest-neighbor connectivity \cite{Bravyi_2010}. To avoid this fundamental challenge, quantum low-density parity-check (LDPC) codes \cite{Breuckmann_2021, HGP_codes, FB_codes, LP_codes, BP_codes, panteleev2022, dinur2022, leverrier2022} have been developed to significantly reduce the resource overheads for error correction \cite{gottesman2014, Fawzi_2018}. To avoid the constraints of Ref.~\cite{Bravyi_2010}, these codes will necessarily require long-range connectivity \cite{Baspin_2022, baspin2023}. Platforms with such capabilities may be able to significantly reduce the overhead for quantum error correction, and a number of recent proposals have been made for how to incorporate limited long-range connectivity into hardware \cite{Bravyi:2023qpn,Xu_2024, LRESC,Viszlai:2023ykh} to ameliorate the constraints of spatial locality.

Thus far, we are not aware of any explicit implementation of a nonlocal LDPC code in any quantum hardware.  Other recent demonstrations of break-even operations we are aware of include the use of a small 7-qubit Steane code~\cite{RyanAnderson2022, Bluvstein_2023} and the 12-qubit carbon code~\cite{silva2024}. While such small codes already improve performance, to avoid large resource overheads in a scalable way, it is crucial to use a nonlocal code which both protects logical qubits in a more complex manner, and is also amenable to quantum hardware.

This work provides a proof-of-principle demonstration that nonlocal codes can improve the performance of present-day quantum computers.  We have implemented a $\llbracket 25,4,3\rrbracket$ quantum LDPC code (encoding 4 logical qubits in 25 data qubits, with the smallest logical operator acting on 3 data qubits) on Quantinuum's H2 trapped-ion quantum computer \cite{Moses2023}, which was configured with 32 physical qubits at the time of the experiments. This encoding enables us to prepare an error-corrected, logical 4-qubit entangled GHZ state beyond the break-even point, with mild postselection overhead, for the first time.  Using the existing hardware, it would not have been possible to attempt such a demonstration, even in principle, by using four surface code patches (with 36 total data qubits) to encode the logical qubits.  This work highlights the importance of choosing good hardware-suited codes with efficient logical gates.  The long-range-enhanced surface code \cite{LRESC} underlying our methods will be scalable to larger trapped-ion QCCD processors \cite{Kielpinski_2002,Home2009,Pino2020,Kaushal2020} as well as to neutral atom arrays \cite{Saffman2016, Browaeys2020, Kaufman_2021, Evered_2023, Bluvstein_2023} for the foreseeable future.

%%%%%%%%%%%%%%%%%%%%%%%%%%%%%%%%%%%%%%%%%%%%%%%%%%%%%%%%%%%%%%%%%

\emph{Quantum encoding.}---
The quantum LDPC code we use is a Calderbank-Shor-Steane (CSS) stabilizer code \cite{gottesman1997, Calderbank_1996, Steane_1996}, which is defined by a stabilizer group of Pauli strings that mutually commute and act trivially on logical codewords.  In a CSS code, these Pauli strings are generated by products of strictly $X$ or $Z$ operators acting on $n$ qubits.  Given the Pauli (anti)commutation relations, it is helpful to organize data about the stabilizer group in a pair of matrices $H_{X,Z}$ with $\mathbb{F}_2=\lbrace0,1\rbrace$ coefficients.  Each row of the matrix $H_X$, which in turn has $n$ columns, fixes one of the stabilizers of the code to be a product of Pauli $X$s on all qubits with a 1 in the corresponding column; $H_Z$ is defined similarly.  To be a valid code, we require $X$ and $Z$ stabilizers to commute, which means $H_XH_Z^{\mathsf{T}}=0$ (here multiplication/addition are mod 2).

A convenient way to build a quantum code is the hypergraph product \cite{HGP_codes} of a classical linear code with itself. For an input $m\times n$ classical parity-check matrix $H$, the resulting CSS stabilizer matrices take the form
\begin{subequations}\label{eq:HGP matrices}
\begin{align}
    H^{\vps}_X &= \big(\, H \otimes \ident_{n} \;\;|\;\; \ident_{m} \otimes H^{\transpose} \,\big) \\
    H^{\vps}_Z &= \big(\, \ident_{n} \otimes H^{} \;\;|\;\; H^{\transpose} \otimes \ident_{m} \,\big) \, .
\end{align}
\end{subequations}
If the classical input code has parameters $[n,k,d]$ with $m=n-k$ parity checks (meaning that there are $n$ classical bits encoding $k$ logicals, with code distance $d$), then so long as the classical code has no redundant parity checks, the resulting quantum CSS hypergraph product code has parameters  $\llbracket n^2+(n-k)^2, k^2, d \rrbracket$. The geometrical layout of the quantum code can be understood as a Cartesian graph product between those of its classical input codes. While the hypergraph product can produce codes with constant rate ($k\propto n$), the distance scaling is sublinear ($d\lesssim\sqrt{n}$). Nonetheless, this distance is preserved under syndrome extraction techniques with circuit-level noise based on ancilla measurement \cite{manes2023}, a feature not supported by other quantum codes such as the Steane code \cite{Steane_1996}.

\begin{figure}[t]
    \centering
    \includegraphics[width=0.49\textwidth]{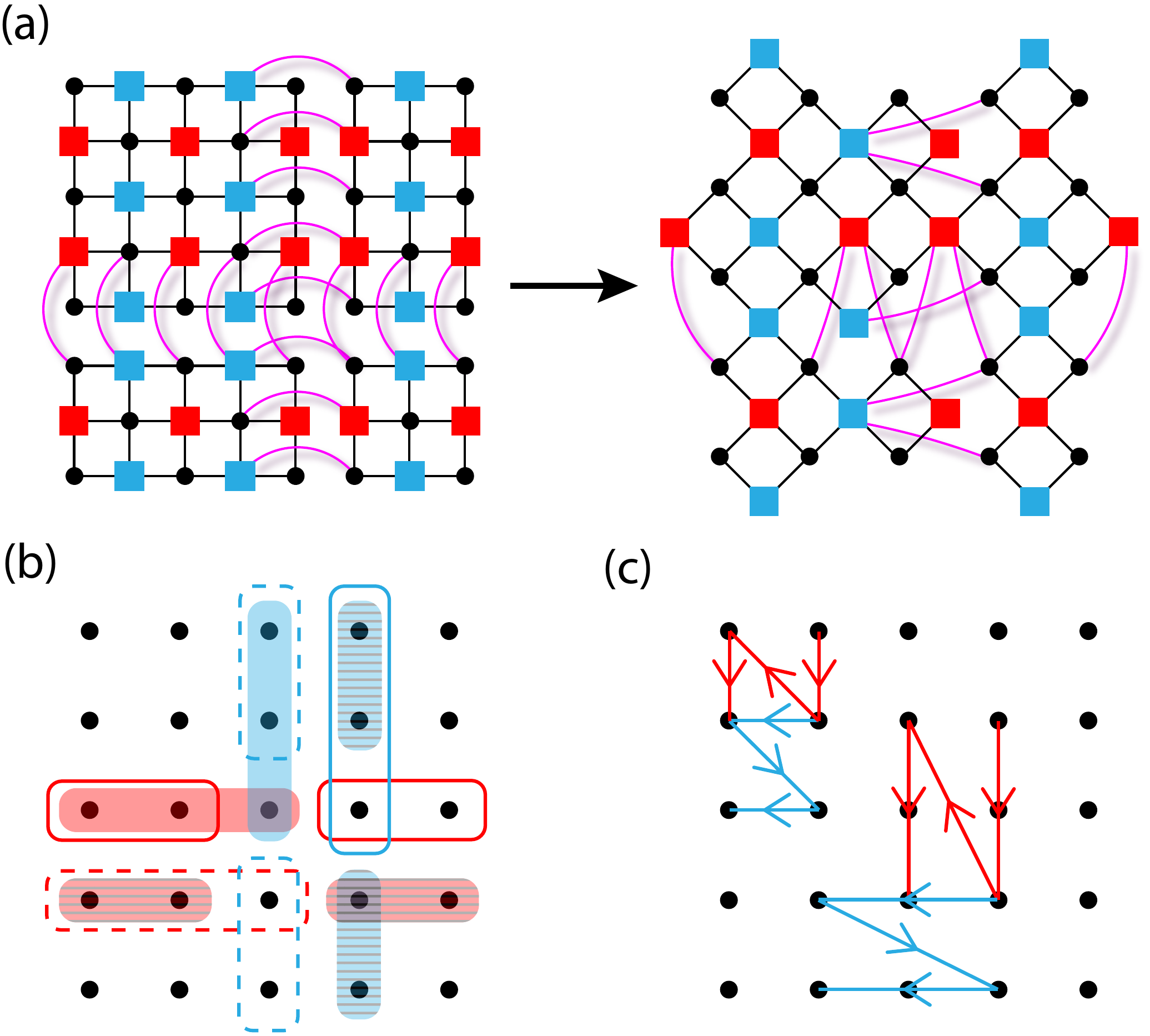}
    \caption{The code construction is illustrated. Circles and red (blue) squares denote qubits and $X$($Z$)-type stabilizers respectively, with edges representing their connectivity. Magenta lines denote long-range interactions. (a) A quantum Tanner transformation reduces the number of data qubits from 34 to 25. (b) The support of the logical $X$ (red) and $Z$ (blue) operators for logical qubits 1-4 are drawn with shaded boxes, solid lines, dashed lines, and striped boxes respectively. (c) For distance-preserving syndrome extraction, the CNOT schedule should follow the direction of the arrows; note that not all checks are shown.}
    \label{fig:code construction}
\end{figure}

We now summarize the construction of our quantum LDPC code, with a graphical depiction in Fig. \ref{fig:code construction}. More details are provided in Appendix \ref{sec:code details}. For the classical input code, we first pick a $[3,2,2]$ parity code consisting of a single, global parity check on all three data bits. We then concatenate two of the data bits with a $[2,1,2]$ repetition code to obtain a $[5,2,3]$ code. We then take the hypergraph product of this $[5,2,3]$ code with itself to obtain a $\llbracket 34,4,3 \rrbracket$ CSS code. Finally, we perform a quantum Tanner transformation \cite{Leverrier_2024} to reduce the number of data qubits from 34 to 25, while preserving both $k$ and $d$. We note that the quantum Tanner transformation does not preserve the generic circuit-level distance property mentioned earlier; however, we can still mitigate the effects of correlated errors by choosing a clever pattern for syndrome extraction like that for the rotated surface code \cite{Tomita_2014}. The order of CNOTs is carefully chosen so that all correlated ancilla (hook) errors end up ``against the grain" of the logical operators, and so $d=3$ errors on the ancilla qubits are necessary to incur an undetectable logical error. This code is a minimal example of a more general family of long-range-enhanced surface codes \cite{LRESC}, which possess the fewest nonlocal stabilizers necessary to avoid the constraints of locality \cite{Bravyi_2010}, while adding O(1) logical qubits.

\begin{figure}[t]
    \centering
    \includegraphics[width=0.49\textwidth]{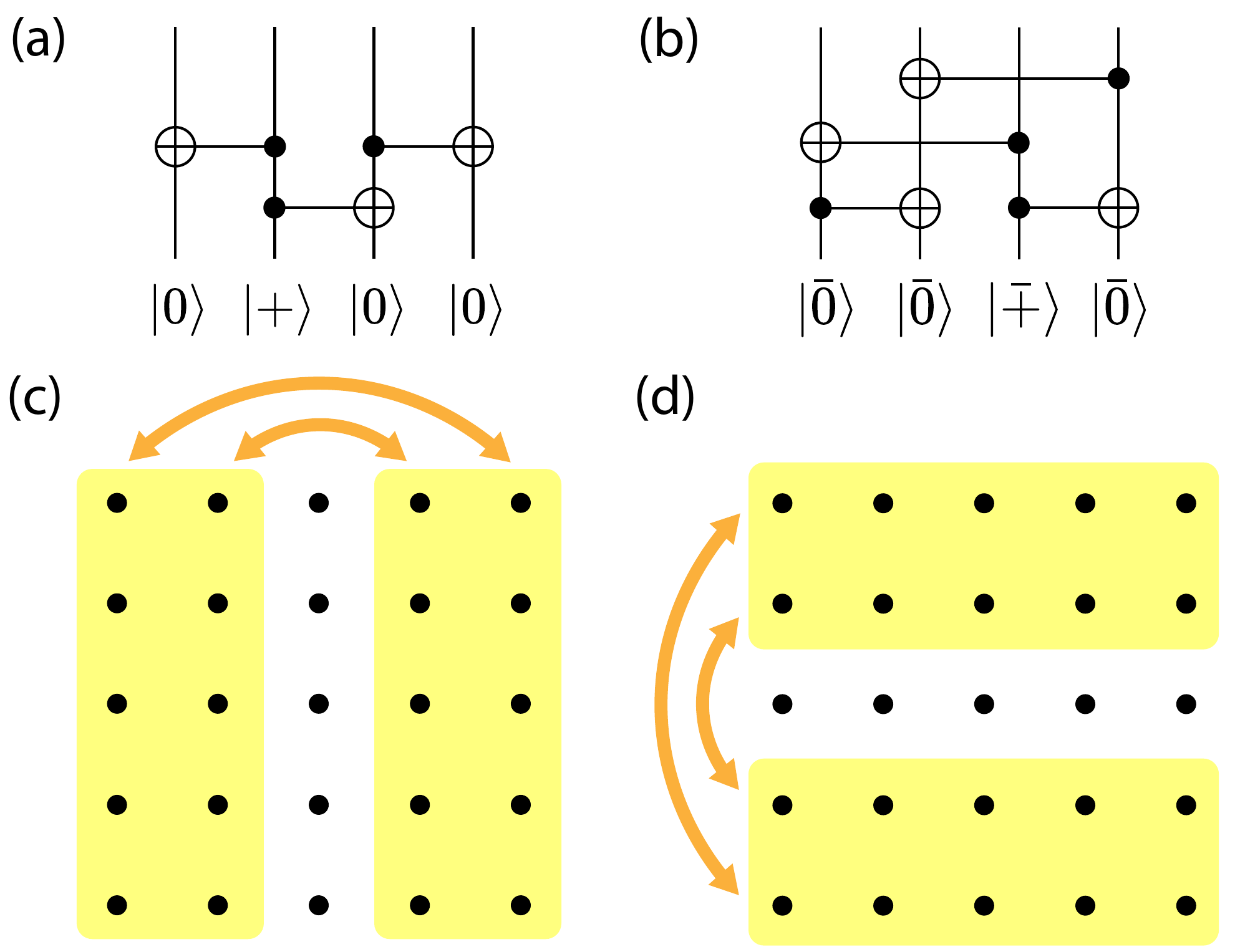}
    \caption{The procedures for preparing the physical and logical 4-qubit GHZ states are illustrated. (a) The quantum circuit which prepares the physical GHZ state. (b) The circuit which prepares the logical GHZ state. (c-d) A combination of fold swaps implements the above logical circuit while preserving the stabilizer group.}
    \label{fig:GHZ circuit}
\end{figure}

The standard protocol to prepare the GHZ state on 4 qubits is to initialize the qubits in $\ket{+000}$ and then apply CNOTs between the $\ket{+}$ qubit (control) and the rest (target); here $Z|0\rangle = |0\rangle$ and $X|+\rangle = |+\rangle$. Since our code is CSS and has a distance-preserving syndrome extraction circuit, we can perform a transversal initialization into the logical $\ket{\bar{0}\bar{0}\bar{0}\bar{0}}$ state by initializing all data qubits in $\ket{0}^{\otimes n}$ and then measuring all $X$-type stabilizers. The stabilizer measurement can be performed by initializing ancillas in the $\ket{+}$ state, performing CNOT gates between the ancillas (control) and data qubits (target) according to the pattern in Fig. \ref{fig:code construction}c, and finally measuring all ancillas in the $X$ basis. We then prepare the third logical qubit in the $\ket{\bar{+}}$ state by measuring one of its logical $\bar{X}_3$ operators three times and postselecting on the agreement of all three outcomes in order to mitigate the effects of readout error. Approximately 98\% of runs had agreement of all 3 outcomes. Assuming postselection onto this 98\%, a logical $\bar{Z}_3$ is then applied (offline) if the agreed outcome is $-1$. After preparing the logical $\ket{\bar{0}\bar{0}\bar{+}\bar{0}}$ state, we perform the desired logical CNOT gates by permuting the data qubits -- a feature inherited from the $[3,2,2]$ parity code. This relabeling step can be done in software in our particular experiment. In particular, two combinations of ``fold-swaps" are sufficient: see Fig. \ref{fig:GHZ circuit}. Finally, we perform a transversal measurement of all logical qubits in either the $X$ or $Z$ basis by measuring all data qubits in that basis. This transversal measurement is then used to reconstruct a final syndrome which is not subject to circuit-level noise and can be used to correct errors. All error correction is performed ``in software" by tracking the Pauli frame: the final logical measurement outcomes are modified according to a decoding algorithm. We use belief-propagation with ordered statistics decoding (BP+OSD) \cite{roffe_2020, BPOSD_2022}, specifically the ``minimum-sum" (10 iterations) and ``combination sweep" (search depth 14) strategies of BP and OS respectively, to infer physical errors based on the syndrome measurement outcomes.

We now summarize the fault tolerance (or lack thereof) of each of the aforementioned steps in our logical GHZ protocol; see Appendix \ref{sec:FT analysis} for more detail. As previously shown, our syndrome extraction circuit is distance-preserving and so we do not need to worry about ancilla hook errors. However, since we only perform a single round of syndrome extraction, one may worry about the effects of ancilla measurement errors. For example, in a large surface code, a single flipped check in the middle of the code may cause a matching decoder to pair it to the boundary, resulting in a large error string. Fortunately, for our distance three code, any single flipped $X$-check can be caused by a single-qubit $Z$ error, and so we do not suffer from this type of error propagation. For the logical $\bar{X}_3$ measurement, a single $Z$ error on a data qubit in the support of $\bar{X}_3$ prior to the measurements can spoil the result, and so this step is not fault tolerant. We note that a fault-tolerant procedure of measuring $\bar{X}_3$ can be done using generalized lattice surgery techniques \cite{Cohen_2022} combined with multiple rounds of syndrome measurements, though this would add considerable overhead (and requires more physical qubits than were available in H2 at the time) and so we leave it to future work. The data qubit permutations to implement the fold-swaps can be achieved by a simple relabeling in software and so is fault tolerant by definition. Lastly, the final measurement of all data qubits is transversal and so is also fault tolerant.

%%%%%%%%%%%%%%%%%%%%%%%%%%%%%%%%%%%%%%%%%%%%%%%%%%%%%%%%%%%%%%%%%

\emph{Experimental results.}--- 
The experiments described in this work used the H2 trapped-ion processor described in Ref.~\cite{Moses2023}. The H2 processor was configured with 32 trapped-ion qubits, all of which can be directly gated with each other using ion-transport operations to pair any two-qubits in a single potential well~\textemdash~a crucial feature for our $\llbracket 25,4,3\rrbracket$ code. We used 25 physical qubits as data qubits and 6 as ancilla qubits for syndrome extraction. Physical-level entangling operations are implemented using a M{\o}lmer-S{\o}rensen interaction \cite{Molmer1999} driven by lasers directed to the four gating regions. Likewise, physical level single-qubit gates, reset, and measurement all take place in these same regions using additional lasers systems. 
%At the time of these experiments, the typical error rates in H2 were as follows: $<2\times10^{-3}$ two-qubit gate error, $\approx 3\times10^{-5}$ single-qubit gate error, and $\approx 2\times10^{-3}$ state preparation and measurement (SPAM) error.

For the physical GHZ experiment, we run the circuit in Fig. \ref{fig:GHZ circuit}a for 10,000 shots, measuring all physical qubits in the $Z$ basis for 5000 shots and in the $X$ basis for the other half. For the logical GHZ experiment, we run the circuit in Fig. \ref{fig:GHZ circuit}b for 5000 shots, with 2500 shots each dedicated to logical $\bar{X}$ and $\bar{Z}$ readout. Uncertainties in the data are computed using the standard error $\sigma = \sqrt{p(1-p)/N}$, where $N$ is the number of shots. For a perfect GHZ state, we expect all $Z$ measurement outcomes to agree and the product of all $X$ measurement outcomes to have even parity. The mismatch rates in the experimental data from these theoretical predictions are tabulated in Table \ref{tab:exp results}. In addition to the error-corrected logical results, we also give the uncorrected logical results which are obtained by simply ignoring the syndrome values (but still postselecting on the logical $\bar{X}_3$ measurement). For completeness, we also ran classical, circuit-level Clifford simulations using Quantinuum's H2-1 emulator which closely resembles the hardware noise. For the classical simulations, we ran 10,000 shots for physical GHZ ($Z$), 5000 shots for physical GHZ ($X$), 5000 shots for logical GHZ ($Z$), and 5000 shots for logical GHZ ($X$). The emulator noise model estimated the following error probabilities: single-qubit gate ( $\approx 3\times 10^{-5}$), two-qubit gate ($2\times 10^{-3}$), and state preparation and measurement, or SPAM ($\approx 2\times 10^{-3}$); further details can be found in \cite{Moses2023}. Simulations slightly overestimated the actual error in the experiment as the machine performance is continually improved.
%over time.

\begin{table}[t]
\renewcommand{\arraystretch}{1.3}
\begin{tabular}{cccc}
\hline
\multicolumn{4}{c}{\textbf{Circuit parameters}}                         \\ \hline
\hline
GHZ experiment     & \;data qubits\; & \;anc. qubits\; & \;2q gates\; \\
\hline
Physical           & 4 & 0 & 3          \\
Logical            & 25 & 6 & 47             \\
\hline
\end{tabular} \\
\vspace{0.5em}
\begin{tabular}{ccc}
\hline
\multicolumn{3}{c}{\textbf{Experimental data}}                         \\ \hline
\hline
GHZ experiment     & \;$\bar{z}$ mismatch (\%)\; & \;$\bar{x}$ mismatch (\%)\; \\
\hline
Physical           & $1.3 \pm 0.2$           & $0.9 \pm 0.1$           \\
Logical (No QEC)   & $5.7 \pm 0.5$           & $2.9 \pm 0.3$           \\
Logical (with QEC) & $0.3 \pm 0.1$           & $0.2 \pm 0.1$           \\
\hline
\end{tabular} \\
\vspace{0.5em}
\begin{tabular}{ccc}
\hline
\multicolumn{3}{c}{\textbf{Classical simulation data}}                         \\ \hline
\hline
GHZ experiment     & \;$\bar{z}$ mismatch (\%)\; & \;$\bar{x}$ mismatch (\%)\; \\
\hline
Physical           & $1.2 \pm 0.1$           & $1.1 \pm 0.1$           \\
Logical (No QEC)   & $5.9 \pm 0.3$           & $3.6 \pm 0.3$           \\
Logical (with QEC) & $0.6 \pm 0.1$           & $0.4 \pm 0.1$           \\
\hline
\end{tabular}
\caption{Fidelity of GHZ state preparation in experiment and classical simulation. The second column displays the percentage of runs when the four $Z$ measurement outcomes disagreed. The third column displays the percentage of runs when the product of the four $X$ measurements was $-1$. Standard errors for the data are also shown.}
\label{tab:exp results}
\end{table}

Since the GHZ state is a stabilizer state with generators $\{ Z_1Z_2, Z_2Z_3, Z_3Z_4, X_1X_2X_3X_4\}$, we can compute simple upper and lower bounds on the GHZ fidelity of our experimental state $\rho$. Because $\ket{\rm GHZ}$ is a pure state, the fidelity \cite{Jozsa_1994} takes the form
\begin{align}\label{eq:fidelity}
    F_{\rm GHZ}(\rho) = \bra{\rm GHZ} \rho \ket{\rm GHZ} \equiv \expval{\rho}_{\rm GHZ} \, .
\end{align}
We now compute an upper bound for the experimental fidelities. First, notice that the two projectors $P^\pm_z$ onto whether the $Z$ measurement outcomes agree ($P^{+}_z=(\ident+Z_1Z_2)(\ident+Z_2Z_3)(\ident+Z_3Z_4)/8$) or disagree ($P^{-}_z=\ident-P^{+}_z$) form a projection-valued measure that splits the Hilbert space into two orthogonal subspaces. Using the fact that $\expval{P^{+}_z}_{\rm GHZ} = 1$, i.e. the GHZ state strictly lives in the $P^{+}_z$ subspace, we can upper bound the fidelity \eqref{eq:fidelity} by
\begin{align}
    F_{\rm GHZ}(\rho) \leq \trace\big(P^{+}_z\rho\big) \, ,
\end{align}
which corresponds to the second column in Table \ref{tab:exp results}. For the lower bound, observe that the GHZ state also strictly resides in the subspace of $P^{+}_x \equiv (\ident+X_1X_2X_3X_4)/2$ such that its density matrix takes the form $\ket{\mathrm{GHZ}}\bra{\mathrm{GHZ}} = P^{+}_z P^{+}_x$. We can then use the union bound from probability theory to upper bound the probabilities of being orthogonal to GHZ, which provides a lower bound to the fidelity \eqref{eq:fidelity}:
\begin{align}\label{eq:fidelity lower bound}
    F_{\rm GHZ}(\rho) \geq 1 - \trace\big(P^{-}_z \rho\big) - \trace\big(P^{-}_x \rho\big) \, .
\end{align}
The first average corresponds to the fraction of runs where the $Z$ outcomes disagree and is given by the second column in Table \ref{tab:exp results}. The contribution from the $XXXX$ generator is given by the third column. For the physical and (error-corrected) logical states $\rho, \bar{\rho}$, we hence obtain fidelity bounds of
\begin{subequations}
\begin{align}
    97.8 \pm 0.2\% &\leq F_{\rm GHZ}(\rho) \leq 98.7 \pm 0.2\%  \\
    99.5 \pm 0.15\% &\leq F_{\rm GHZ}(\bar{\rho}) \leq 99.7 \pm 0.1\% \, .
\end{align}
\end{subequations}
The statistical significance between the physical upper bound and the logical lower bound is $\approx 3.6\sigma$, assuming statistically independent uncertainties.

%%%%%%%%%%%%%%%%%%%%%%%%%%%%%%%%%%%%%%%%%%%%%%%%%%%%%%%%%%%%%%%%

\emph{Outlook.}---
We have prepared a logical GHZ state on four qubits with a higher fidelity than its physical counterpart. The key ingredient is a quantum LDPC code whose long-range interactions enable (i) a compact encoding of logical qubits and (ii) an efficient implementation of specific logical circuits via qubit permutations.  Having demonstrated the ability to prepare specific quantum entangled logical states beyond break-even, an immediate near-term experimental goal will be to use our encoding (or others) to demonstrate break-even on larger circuits involving multiple logical qubits, eventually including circuits with non-Clifford gates. We also emphasize that while our code is able to efficiently perform parallel CNOT gates between the four logical qubits, it cannot perform arbitrary targeted single-qubit and two-qubit gates, and so achieving universality would require additional gadgets. For example, Appendix \ref{sec:generalized GHZ} describes a simple generalization of our experiment that would prepare larger GHZ states based on similar codes and logical gadgets, and explains how such operations can be made fault-tolerant with increasing $\llbracket n,k,d\rrbracket$.

Our work highlights an advantage of using quantum LDPC codes in existing hardware over conventional topological codes like the surface and color codes. Of course, a major feature of the topological codes is that they are equipped with transversal implementations of the whole Clifford group. A major challenge in the field is to construct nonlocal quantum LDPC codes with hardware-efficient logical gates; some progress using hypergraph product codes has been previously made \cite{Krishna_2021, Quintavalle_2023}. We hope that this work stimulates further progress in this direction, especially since it is already known that LDPC codes can already reduce the overhead for fault-tolerant quantum computation when functioning just as a memory block \cite{Cohen_2022}. On a separate front, nonlocal codes derived from concatenating quantum Hamming codes also show promise for low-overhead fault tolerance \cite{Yamasaki_2024, yoshida2024}. Quantum Hamming codes in particular have the capability of implementing the entire logical Clifford group with low overhead \cite{Chao_2018}. However, since these codes are not LDPC, they require more complicated gadgets for state initialization and error correction \cite{Chao_2018_two}. A worthwhile future direction is to perform a detailed cost-benefit analysis between LDPC codes and concatenated ones on various hardware.

%%%%%%%%%%%%%%%%%%%%%%%%%%%%%%%%%%%%%%%%%%%%%%%%%%%%%%%%%%%%%%%%

\emph{Acknowledgements.}---  This work was supported in part by the Alfred P. Sloan Foundation under Grant FG-2020-13795 (AL), by the Air Force Office of Scientific Research under Grant FA9550-24-1-0120 (YH, AL), and by the Office of Naval Research via Grant N00014-23-1-2533 (YH, AL).
Additionally, we thank the hardware team at Quantinuum for making these experiments possible.

%%%%%%%%%%%%%%%%%%%%%%%%%%%%%%%%%%%%%%%%%%%%%%%%%%%%%%%%%%%%%%%%%%%%%%%%%%
%%%%%%%%%%%%%%%%%%%%%%%%%%%%%%%%%%%%%%%%%%%%%%%%%%%%%%%%%%%%%%%%%%%%%%%%%%
%%%%%%%%%%%%%%%%%%%%%%%%%%%%%%%%%%%%%%%%%%%%%%%%%%%%%%%%%%%%%%%%%%%%%%%%%%

\clearpage
\onecolumngrid
\appendix
\renewcommand{\thesubsection}{\thesection.\arabic{subsection}}

\section{Code details}\label{sec:code details}
We first provide more details and intuition into the code analyzed in the main text.

\subsection{Hypergraph product and logical operators}
\label{sec:HGP and logicals}

\begin{figure}[t]
    \centering
    \includegraphics[width=0.9\linewidth]{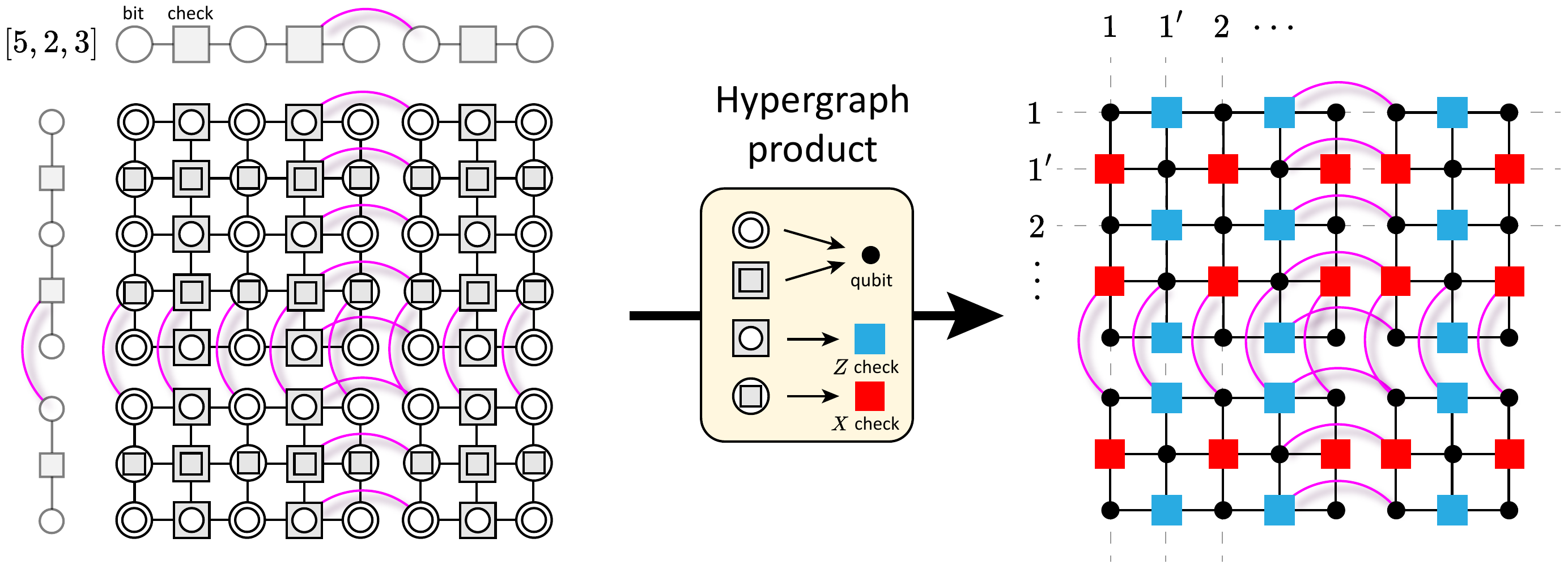}
    \caption{The hypergraph product of the $[5,2,3]$ parent code is illustrated. A Euclidean graph product of the $[5,2,3]$ Tanner graph with itself produces a new graph with four types of vertices, and the hypergraph product transforms this product graph into a quantum CSS Tanner graph. The bit-bit and check-check qubits live on the primary and secondary lattices respectively, which are labeled by unprimed and primed indices. Edges denote the connectivity between qubits and checks.}
    \label{fig:HGP detailed}
\end{figure}

We begin by discussing the non-Tanner-transformed hypergraph product code.  The parity-check $H$ and generator $G$ matrices of the $[5,2,3]$ parent code input into the hypergraph product are given by
\begin{align}\label{eq:[5,2,3] H and G}
    H = \left(\hspace{-0.25em}\begin{array}{ccccc} 1&1&0&0&0 \\ 0&1&1&1&0 \\ 0&0&0&1&1 \end{array}\hspace{-0.25em}\right) \quad,\quad G = \left(\hspace{-0.25em}\begin{array}{cc|cc|c} 1&1&1&0&0 \\ 1&1&0&1&1 \end{array}\hspace{-0.25em}\right) \, ,
\end{align}
where we have put $G$ in reduced-row echelon form and explicitly highlighted its $2\times 2$ identity block. Note that $HG^\transpose = 0$ over $\mathbb{F}_2$ as required. As depicted in Fig. \ref{fig:HGP detailed}, its associated hypergraph product \cite{HGP_codes} code contains 34 qubits arranged in a $5\times 5$ primary lattice of ``bit-bit'' qubits and a $3\times 3$ secondary lattice of ``check-check'' qubits. As a convention, we will index the rows and columns of the primary (secondary) lattice with unprimed (primed) indices. $X$-type and $Z$-type checks are then labeled by coordinates $(i',j)$ and $(i,j')$ respectively. For convenience, also write $X_{i,j}$ to denote the Pauli $X$ operator on site $(i,j)$.

The support of logical Pauli $\bar{X}$ and $\bar{Z}$ operators can be read off from the generator matrix $G$ in \eqref{eq:[5,2,3] H and G}. In particular, note that the identity block resides in columns 3 and 4 of $G$. We choose logical $\bar{X}$ ($\bar{Z}$) operators to have support on rows (columns) 3 and 4; the other coordinate is then given by the position of 1s in each row of $G$. Explicitly, the four logical qubits have logical Pauli operators given by
\begin{enumerate}
    \item $\bar{X}_1 = X_{3,1} X_{3,2} X_{3,3}$ , \, $\bar{Z}_1 = Z_{1,3} Z_{2,3} Z_{3,3}$
    \item $\bar{X}_2 = X_{3,1} X_{3,2} X_{3,4} X_{3,5}$ , \, $\bar{Z}_2 = Z_{1,4} Z_{2,4} Z_{3,4}$
    \item $\bar{X}_3 = X_{4,1} X_{4,2} X_{4,3}$ , \, $\bar{Z}_3 = Z_{1,3} Z_{2,3} Z_{4,3} Z_{5,3}$
    \item $\bar{X}_4 = X_{4,1} X_{4,2} X_{4,4} X_{4,5}$ , \, $\bar{Z}_4 = Z_{1,4} Z_{2,4} Z_{4,4} Z_{5,4}$
\end{enumerate}
Note that each pair of logical operators $\bar{X}_i, \bar{Z}_i$ overlap on exactly one qubit, and thus anticommute as required. Importantly, the logicals operators have a directional ``grain'': $\bar{X}$ operators traverse horizontally while $\bar{Z}$ operators traverse vertically (see Fig. 1b in the main paper), just like in the surface code.

\subsection{Quantum Tanner transformation}
\label{sec:QTT}

\begin{figure}[t]
    \centering
    \includegraphics[width=0.66\linewidth]{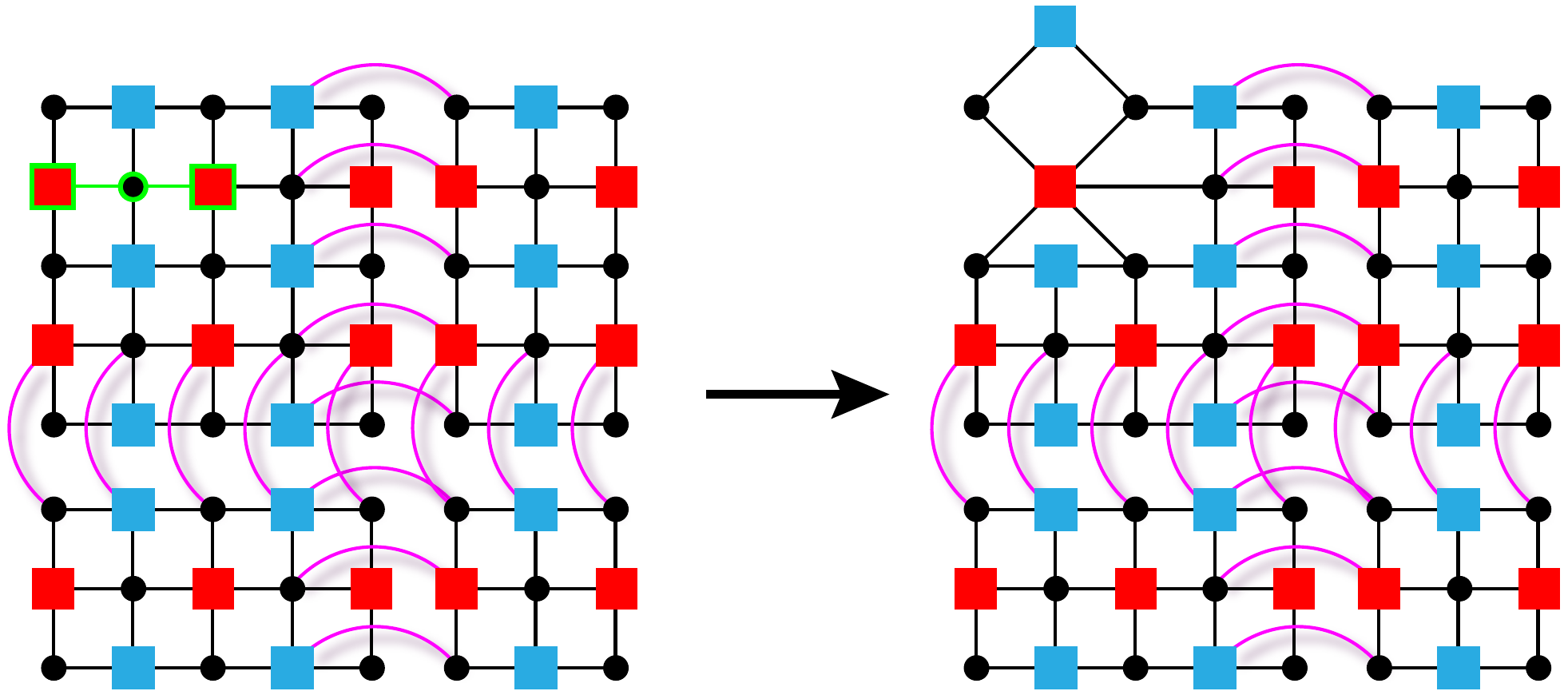}
    \caption{One step of the quantum Tanner transformation is depicted. A secondary qubit is chosen, and its incident $X$ checks (highlighted in green) are rearranged such that the secondary qubit can be discarded.}
    \label{fig:QTT step}
\end{figure}

The idea behind the quantum Tanner transformation (QTT) \cite{Leverrier_2024} is to rearrange the stabilizer generators (checks) such that the secondary lattice (check-check qubits) can be discarded, thereby decreasing the number of data qubits while maintaining the number of logical qubits and the code distance. We can perform the QTT iteratively as follows:
\begin{enumerate}
    \item Choose a qubit on the secondary lattice indexed by $(i',j')$ and a Pauli type ($X$ or $Z$).
    \item Examine the incident checks to $(i',j')$ of the Pauli type chosen in step 1. Select one check and replace every other incident check (of the same type) with the combination of itself and the chosen check. Discard the chosen check and the qubit $(i',j')$.
\end{enumerate}
The above procedure is iterated 9 times for our $\llbracket 34,4,3 \rrbracket$ hypergraph product code since we have 9 secondary qubits; see Fig. \ref{fig:QTT step} for an illustration of one step of the QTT. Note that in step 2, because we are combining neighboring checks, the weight of the checks will generally increase. However, we have the freedom to choose the Pauli type in step 1, and so we will typically alternate between $X$ and $Z$-type checks for each iteration in order to minimize the final check weights. Due to the nature of the graph product, every $X$ and $Z$ check either has no overlap or overlap on exactly two qubits: one on the primary lattice and one on the secondary lattice. By multiplying, say, $X$ checks incident to a secondary qubit according to step 2 above, we construct a new set of local checks which commute with the incident $Z$ checks solely on the primary lattice. We then construct a new lattice by discarding the secondary qubit. Since the checks represent generators of the stabilizer group, their combinations also form valid generators. Furthermore, since the new checks commute on the restricted sublattice without the secondary qubit, the stabilizer group acts equivalently on the new lattice.

For the post-QTT $\llbracket 25,4,3 \rrbracket$ quantum code, we can choose the same logical Pauli operators as in Sec. \ref{sec:HGP and logicals} since they were supported entirely on the primary lattice. Since the stabilizer group remained unchanged on the primary lattice during the QTT, the code distance is the same as before.   For concreteness, we also present formulas for the stabilizer group generators $\mathcal{S}$ of the final Tanner transformed code, which are depicted graphically in Fig. \ref{fig:code construction} and in Fig. \ref{fig:syndrome measurement schedule}a:
\begin{align}
    \mathcal{S} &= \lbrace Z_{1,1}Z_{1,2}, \; Z_{1,4}Z_{1,5}, \; Z_{5,1}Z_{5,2}, \; Z_{5,4}Z_{5,5}, \; Z_{2,1}Z_{2,2}Z_{3,1}Z_{3,2}, \;  Z_{2,4}Z_{2,5}Z_{3,4}Z_{3,5}, \; Z_{3,1}Z_{3,2}Z_{4,1}Z_{4,2}, \notag \\
    &\;\;\; Z_{3,4}Z_{3,5}Z_{4,4}Z_{4,5}, \; Z_{3,2},Z_{3,3},Z_{3,4},  \; Z_{1,2}Z_{1,3}Z_{1,4}Z_{2,2}Z_{2,3}Z_{2,4}, \; Z_{4,2}Z_{4,3}Z_{4,4}Z_{5,2}Z_{5,3}Z_{5,4}, \notag \\
    & \;\;\; X_{3,1}X_{3,2}, \; X_{3,4}X_{3,5}, \; X_{1,1}X_{1,2}X_{2,1}X_{2,2}, \; X_{4,1}X_{4,2}X_{5,1}X_{5,2}, \;\; X_{1,4}X_{1,5}X_{2,4}X_{2,5}, \; X_{4,4}X_{4,5}X_{5,4}X_{5,5}, \notag \\
    & \;\;\; X_{2,1}X_{3,1}X_{4,1}, \; X_{2,5}X_{3,5}X_{4,5}, \; X_{2,2}X_{3,2}X_{4,2}X_{2,3}X_{3,3}X_{4,3},\; X_{2,4}X_{3,4}X_{4,4}X_{2,3}X_{3,3}X_{4,3}  \rbrace . \label{eq:Sgroup}
\end{align}

\subsection{Fold-swap logical gates}

The post-QTT $\llbracket 25,4,3 \rrbracket$ code is depicted in Fig. \ref{fig:syndrome measurement schedule}a. Using the coordinate convention described in Sec. \ref{sec:HGP and logicals}, we now analyze the fold-swap gadgets in detail. The vertical fold-swap consists of swapping columns $\{ (\cdot,1), (\cdot,2) \}$ with columns $\{ (\cdot,5), (\cdot,4) \}$ respectively. Using the logical operators defined in Sec. \ref{sec:HGP and logicals}, the logical $\bar{X}$ operators transform as
\begin{align}\label{eq:vertical swap logical Xs}
    \bar{X}_1 &\longrightarrow X_{3,3} X_{3,4} X_{3,5} = \bar{X}_1 \bar{X}_2  \notag \\
    \bar{X}_2 &\longrightarrow \bar{X}_2  \notag \\
    \bar{X}_3 &\longrightarrow X_{4,3} X_{4,4} X_{4,5} = \bar{X}_3 \bar{X}_4  \notag \\
    \bar{X}_4 &\longrightarrow \bar{X}_4 \, .
\end{align}
The logical $\bar{Z}$ operators transform as
\begin{align}\label{eq:vertical swap logical Zs}
    \bar{Z}_1 &\longrightarrow \bar{Z}_1  \notag \\
    \bar{Z}_2 &\longrightarrow Z_{1,2} Z_{2,2} Z_{3,2} \sim \bar{Z}_1 \bar{Z}_2  \notag \\
    \bar{Z}_3 &\longrightarrow \bar{Z}_3  \notag \\
    \bar{Z}_4 &\longrightarrow Z_{1,2} Z_{2,2} Z_{4,2} Z_{5,2} \sim \bar{Z}_3 \bar{Z}_4 \, ,
\end{align}
where $\sim$ denotes an equivalence under the stabilizer group; e.g. $Z_{1,2} Z_{2,2} Z_{3,2}$ is related to $Z_{1,3} Z_{2,3} Z_{3,3} Z_{1,4} Z_{2,4} Z_{3,4} = \bar{Z}_1 \bar{Z}_2$ by the stabilizer element $Z_{1,2} Z_{2,2} Z_{3,2} Z_{1,3} Z_{2,3} Z_{3,3} Z_{1,4} Z_{2,4} Z_{3,4}$. Examining the logical Pauli transformations in \eqref{eq:vertical swap logical Xs} and \eqref{eq:vertical swap logical Zs}, we deduce that the vertical fold-swap implements $\mathrm{CNOT}_{\bar{1}\rightarrow\bar{2}} \cdot \mathrm{CNOT}_{\bar{3}\rightarrow\bar{4}}$ at the logical level.

We analyze the horizontal fold-swap similarly. The horizontal fold-swap consists of swapping rows $\{ (1,\cdot), (2,\cdot) \}$ with rows $\{ (5,\cdot), (4,\cdot) \}$ respectively. The logical $\bar{X}$ operators transform as
\begin{align}\label{eq:horizontal swap logical Xs}
    \bar{X}_1 &\longrightarrow \bar{X}_1  \notag \\
    \bar{X}_2 &\longrightarrow \bar{X}_2  \notag \\
    \bar{X}_3 &\longrightarrow X_{2,1} X_{2,2} X_{2,3} \sim \bar{X}_1 \bar{X}_3  \notag \\
    \bar{X}_4 &\longrightarrow X_{2,1} X_{2,2} X_{2,4} X_{2,5} \sim \bar{X}_2 \bar{X}_4  \, .
\end{align}
The logical $\bar{Z}$ operators transform as
\begin{align}\label{eq:horizontal swap logical Zs}
    \bar{Z}_1 &\longrightarrow Z_{3,3} Z_{4,3} Z_{5,3} = \bar{Z}_1 \bar{Z}_3  \notag \\
    \bar{Z}_2 &\longrightarrow Z_{3,4} Z_{4,4} Z_{5,4} = \bar{Z}_2 \bar{Z}_4  \notag \\
    \bar{Z}_3 &\longrightarrow \bar{Z}_3  \notag \\
    \bar{Z}_4 &\longrightarrow \bar{Z}_4 \, .
\end{align}
From \eqref{eq:horizontal swap logical Xs} and \eqref{eq:horizontal swap logical Zs}, we deduce that the horizontal fold-swap implements $\mathrm{CNOT}_{\bar{3}\rightarrow\bar{1}} \cdot \mathrm{CNOT}_{\bar{4}\rightarrow\bar{2}}$ at the logical level.

Lastly, one can directly confirm that the column and row swaps described above are automorphisms of the code, either by staring at Fig. \ref{fig:syndrome measurement schedule}a, or by explicitly checking that $\mathcal{S}$ given in (\ref{eq:Sgroup}) is invariant under these row and column swaps.

%%%%%%%%%%%%%%%%%%%%%%%%%%%%%%%%%%%%%%%%%%%%%%%%%%%%%%%%%%%%%%%%%
%%%%%%%%%%%%%%%%%%%%%%%%%%%%%%%%%%%%%%%%%%%%%%%%%%%%%%%%%%%%%%%%%

\section{Fault tolerance analysis}\label{sec:FT analysis}

In the first part of this section, we analyze the fault tolerance of our syndrome extraction. In the second part, we discuss the non-fault-tolerant aspect of the logical $\bar{X}_3$ measurement step in the GHZ protocol.

\subsection{Hook and measurement errors}

\begin{figure}[t]
    \centering
    \includegraphics[width=0.8\linewidth]{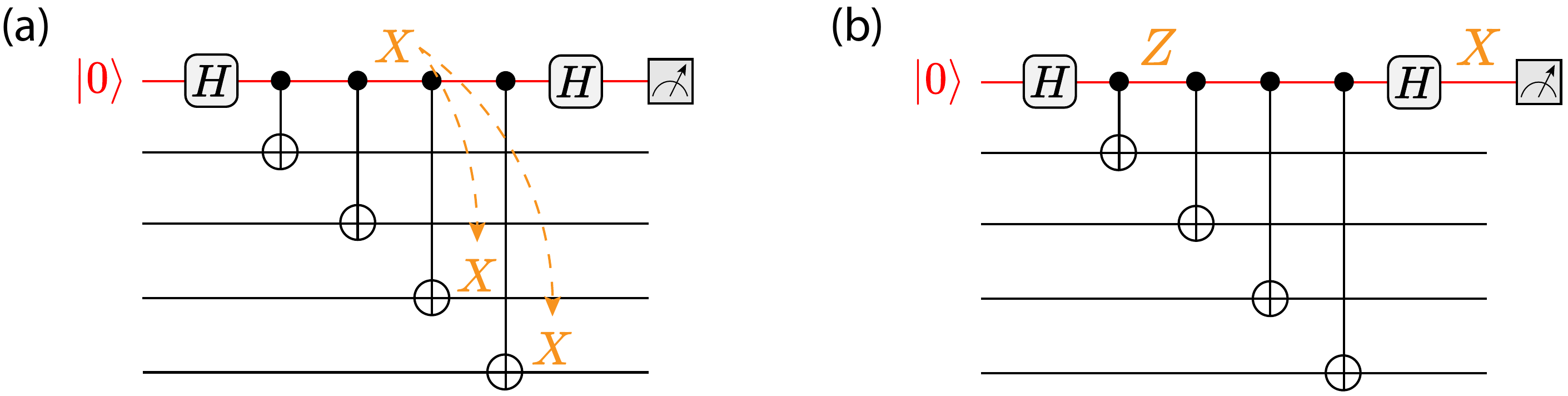}
    \caption{The ancilla measurement gadget is depicted for a weight-4 $X$-check measurement. An ancilla qubit (red) initialized in $\ket{0}$ is introduced, coupled to the data qubits (black), and finally measured in the computational basis. (a) An ancillary $X$ error in the middle of the circuit propagates to two data $X$ errors. (b) An ancillary $Z$ error in the middle of the circuit flips the outcome of the final measurement.}
    \label{fig:hook error circuit}
\end{figure}

To measure the multi-Pauli check operators in the $\llbracket 25,4,3 \rrbracket$ code, we use standard ancilla measurement gadgets as depicted in Fig. \ref{fig:hook error circuit}. Since our code is a CSS code, error correction can be decomposed into and independently analyzed on $X$ and $Z$ errors. There are two kinds of circuit-level errors that we need to worry about. The first is a correlated ``hook'' error between the ancilla and data qubits as depicted in Fig. \ref{fig:hook error circuit}a, and the second is a measurement error as depicted in Fig. \ref{fig:hook error circuit}b. Hook errors are potentially dangerous because a single-qubit error on the ancilla can propagate to a multi-qubit error on the data qubits and potentially overwhelm the error-correcting capability of the code. Measurement errors can also be detrimental because the decoder is essentially fed inaccurate information and can output a wrong correction that may increase the error instead of reduce it. For the hook errors, recall that errors related by an element of the stabilizer group are equivalent. For example, an $X$ error on the ancilla in Fig. \ref{fig:hook error circuit}b before the first CNOT gate propagates to $S_X$ on the data qubits, where $S_X$ is the $X$-check. However, since $S_X$ is an element of the stabilizer group, $S_X \sim \ident$. Thus, all correlated hook errors have effective support within either the first or second half of the ancilla measurement gadget.

\begin{figure}[t]
    \centering
    \includegraphics[width=0.7\linewidth]{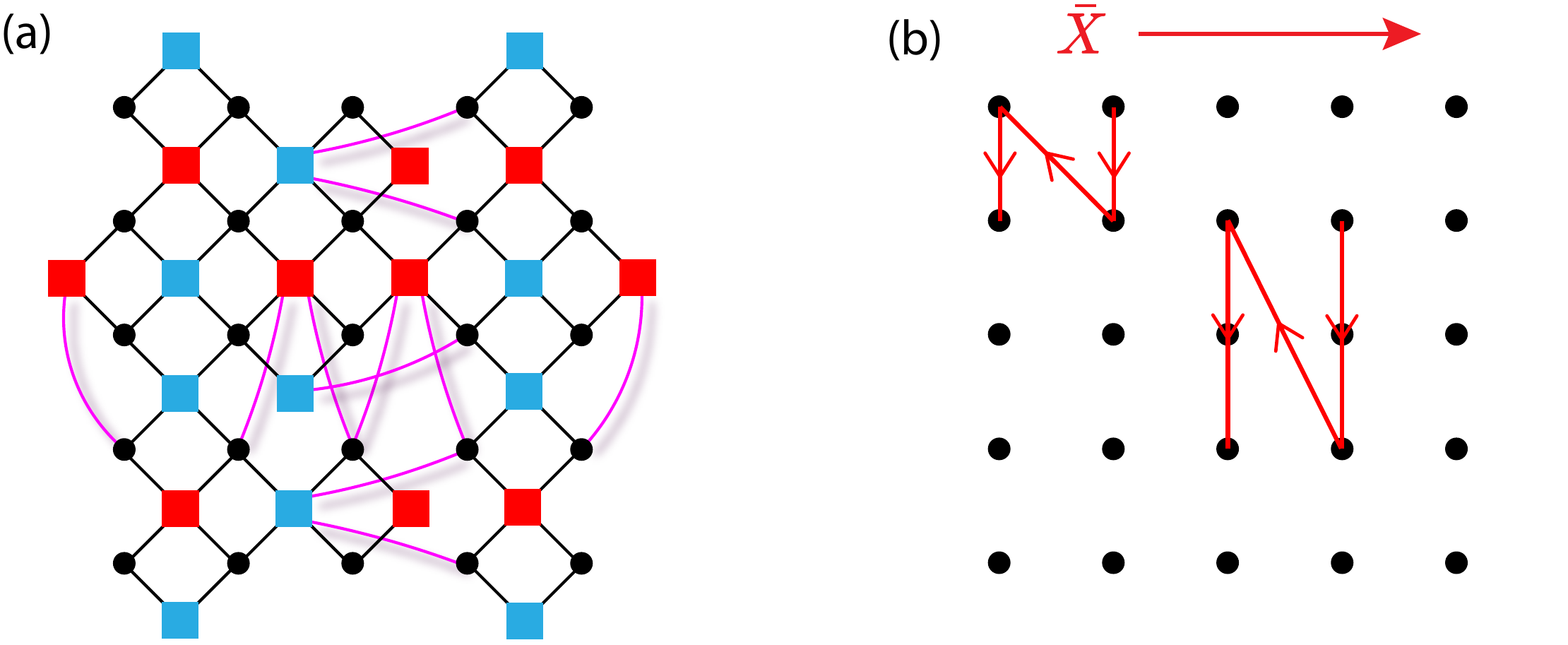}
    \caption{(a) The layout of the checks in the $\llbracket 25,4,3 \rrbracket$ code is shown. (b) The schedule of CNOT gates used to measure the $X$ checks as well as the logical $\bar{X}$ grain direction are shown.}
    \label{fig:syndrome measurement schedule}
\end{figure}

To deal with hook errors, we perform the QTT in a specific manner so that the checks have support on rectangles whose width along the corresponding logical grain is at most 2; see Fig. \ref{fig:syndrome measurement schedule}a. We then schedule the CNOT gates in the ancilla measurement gadget in the zigzag pattern depicted in Fig. \ref{fig:syndrome measurement schedule}b. In this manner, all $X$-type ancilla hook errors have support on vertical strips, and are thus ``against the grain'' of the $\bar{X}$ logical operators. The procedure for $Z$ errors follows analogously but in the horizontal direction. In other words, at least $d=3$ ancilla hook errors are required to complete a logical operation. We call such a measurement schedule \emph{distance preserving}.

In the surface code, measurement errors can be very detrimental. For example, suppose we perform a single round of decoding on a large surface code, and our only error is a measurement error on a check in the middle of the code. A matching decoder will then pair this ``anyon'' to the boundary, implementing a large Pauli error which may become uncorrectable in the next round of decoding. The standard technique to deal with measurement errors is to perform $\Theta(d)$ rounds of error correction and decode over the spacetime syndrome history \cite{Dennis_2002}. Fortunately, there is a special property of the $d=3$ surface code, as well as our $\llbracket 25,4,3 \rrbracket$ code, which permits a single round of syndrome measurement to be fault tolerant to single errors (which is the most we can ask of a $d=3$ code anyway): any single check can be flipped by a single-qubit error. One can quickly verify this property by examining the stabilizer checks in \eqref{eq:Sgroup}. For example, the two weight-6 checks in Fig. \ref{fig:syndrome measurement schedule}a can be individually flipped via $Z$ errors on the qubits at coordinates $(3,2)$ and $(3,4)$ respectively. Thus, any single incorrectly flipped check, through a maximum-likelihood decoder, can only propagate to a single-qubit error on the data qubits.

\subsection{Single logical measurement}

One of the intermediate steps of the GHZ protocol requires applying a targeted logical Hadamard gate to one of the logical qubits. Since all of our logical qubits are initialized in $\ket{\bar{0}}$, we achieve the logical Hadamard by measuring the $\bar{X}$ operator of the desired logical qubit. We perform this measurement by using an ancilla measurement gadget (Fig. \ref{fig:hook error circuit}) to measure $\bar{X}_3$ three times and postselecting on the agreement of the outcomes. This procedure is not fault tolerant because a single $Z$ error on any of the qubits in the support of $\bar{X}_3$ (namely either $X_{4,1}$, $X_{4,2}$ or $X_{4,3}$) prior to the measurement gadget will spoil all three measurement results. In addition, the postselection is not scalable for larger code sizes.

A fault-tolerant and scalable procedure could proceed as follows. First, we initialize an ancillary $d=3$ surface-code in the logical $\ket{\bar{+}}$ state, which requires 9 additional data qubits. We then merge this surface code with our $\llbracket 25,4,3 \rrbracket$ code along the $\bar{X}_3$ operator by measuring appropriate intermediate checks \cite{Cohen_2022}. The outcome of this merge is effectively measuring $\bar{X}_3$: the parity of measurement outcomes from these intermediate checks is the eigenvalue of $\bar{X}_3$. Three rounds of syndrome measurements are then required for fault tolerance against measurement errors. We then return to the original codespace by measuring all 9 appended surface-code qubits in the $Z$ basis.

%%%%%%%%%%%%%%%%%%%%%%%%%%%%%%%%%%%%%%%%%%%%%%%%%%%%%%%%%%%%%%%%%
%%%%%%%%%%%%%%%%%%%%%%%%%%%%%%%%%%%%%%%%%%%%%%%%%%%%%%%%%%%%%%%%%

\section{Generalization to larger GHZ states}
\label{sec:generalized GHZ}

A simple way to generate larger logical GHZ states is to first prepare a smaller logical GHZ state, in say the $\llbracket 25,4,3 \rrbracket$ code, and then initialize additional $\llbracket 25,4,3 \rrbracket$ code blocks in $\ket{\bar{0}}^{\otimes 4}$ and perform transversal CNOT between the GHZ block and the new blocks. However, in the spirit of the $\llbracket 25,4,3 \rrbracket$ code, we can construct bigger code blocks with more logical qubits such that all necessary logical entangling gates can be performed by qubit permutations.

We now describe a generalization of our protocol to prepare a GHZ state on $2(\ell-1)$ logical qubits for $\ell \geq 3$. Instead of the $[3,2,2]$ parity code, we examine the generic case of a $[\ell,\ell-1,2]$ parity code with matrices
\begin{align}\label{eq:general parity code}
    H = \underbrace{\begin{pmatrix} 1 & 1 & \dots & 1 \end{pmatrix}}_\text{$\ell$ times} \quad,\quad G = \begin{pNiceArray}{cw{c}{0.75cm}c|c}[margin]
    \Block{3-3}<\Large>{\ident^{}_{\ell-1}} & & & 1 \\
    & & & \Vdots \\
    & & & 1
    \end{pNiceArray} \, .
\end{align}

\begin{figure}[t]
    \centering
    \includegraphics[width=0.8\linewidth]{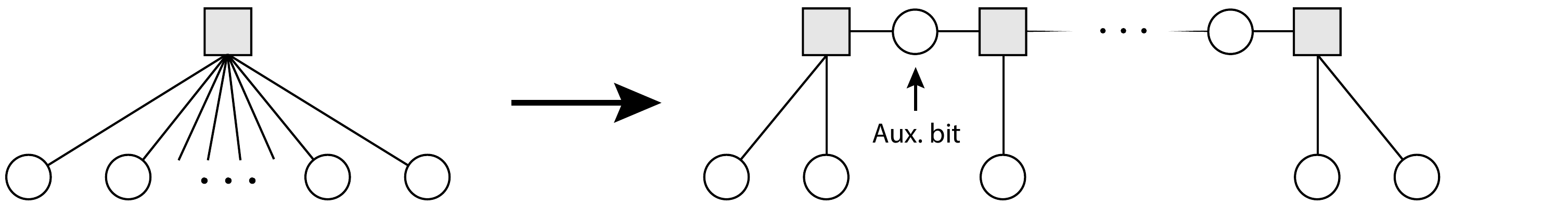}
    \caption{The weight balancing procedure to transform the $[\ell,\ell-1,2]$ parity code into a $[2\ell-3,\ell-1,2]$ LDPC code with $\ell-2$ weight-3 checks is depicted.}
    \label{fig:weight balancing}
\end{figure}

The permutation automorphism group of $H$ is the symmetric group $\mathbf{S}_\ell$ since $H$ is invariant under arbitrary permutations of its columns. Let us analyze the action of $\mathbf{S}_\ell$ on the logical space. Examining $G$, we observe that permutations of the first $\ell-1$ columns, corresponding to data bits, maps to the same permutations of the $\ell-1$ rows, corresponding to logical bits. Swapping the last column of 1s with column $i\neq \ell$ implements the following row transformations on $G$:
\begin{align}\label{eq:general parity swap}
    g_i &\longrightarrow g_i  \notag \\
    g_j &\longrightarrow g_i + g_j \;,\;\; \forall j \neq i \, ,
\end{align}
where $g_j$ denotes the $j$th row of $G$ in \eqref{eq:general parity code}. Unfortunately, the $H$ in \eqref{eq:general parity code} is not LDPC, and so one may worry about fault tolerance for large $\ell$. However, we can perform weight reduction \cite{LRESC} by introducing $\ell-3$ auxiliary bits and $\ell-3$ additional checks so that all checks have weight at most 3: see Fig. \ref{fig:weight balancing} for an illustration. The new $[2\ell-3,\ell-1,2]$ code has a generator matrix in the block form
\begin{align}
    \tilde{G} = \left(\begin{array}{c|c} G & R \end{array}\right) \, ,
\end{align}
for some $(\ell-1) \times (\ell-3)$ binary matrix $R$ whose columns index the auxiliary bits. The weight-reduction procedure reduces the automorphism group to a smaller subgroup, but importantly it preserves \eqref{eq:general parity swap} for $i=\ell-1$ (the rightmost check in Fig. \ref{fig:weight balancing} is invariant under swapping the two rightmost bits).

To increase the code distance beyond 2, we concatenate the $[2\ell-3,\ell-1,2]$ code with a $[c,1,c]$ repetition code to produce a $[(2\ell-3)c,\ell-1,2c]$ code. Our quantum LDPC code is then given by the hypergraph product of the $[(2\ell-3)c,\ell-1,2c]$ code with a concatenated $[3c,2,2c]$ parity code with parameters
\begin{align}
    n &= 6(2\ell-3)c^2 - (7\ell-9)c + 2\ell - 2  \notag \\
    k &= 2(\ell-1)  \notag \\
    d &= 2c \, .
\end{align}

\begin{figure}[t]
    \centering
    \includegraphics[width=0.4\linewidth]{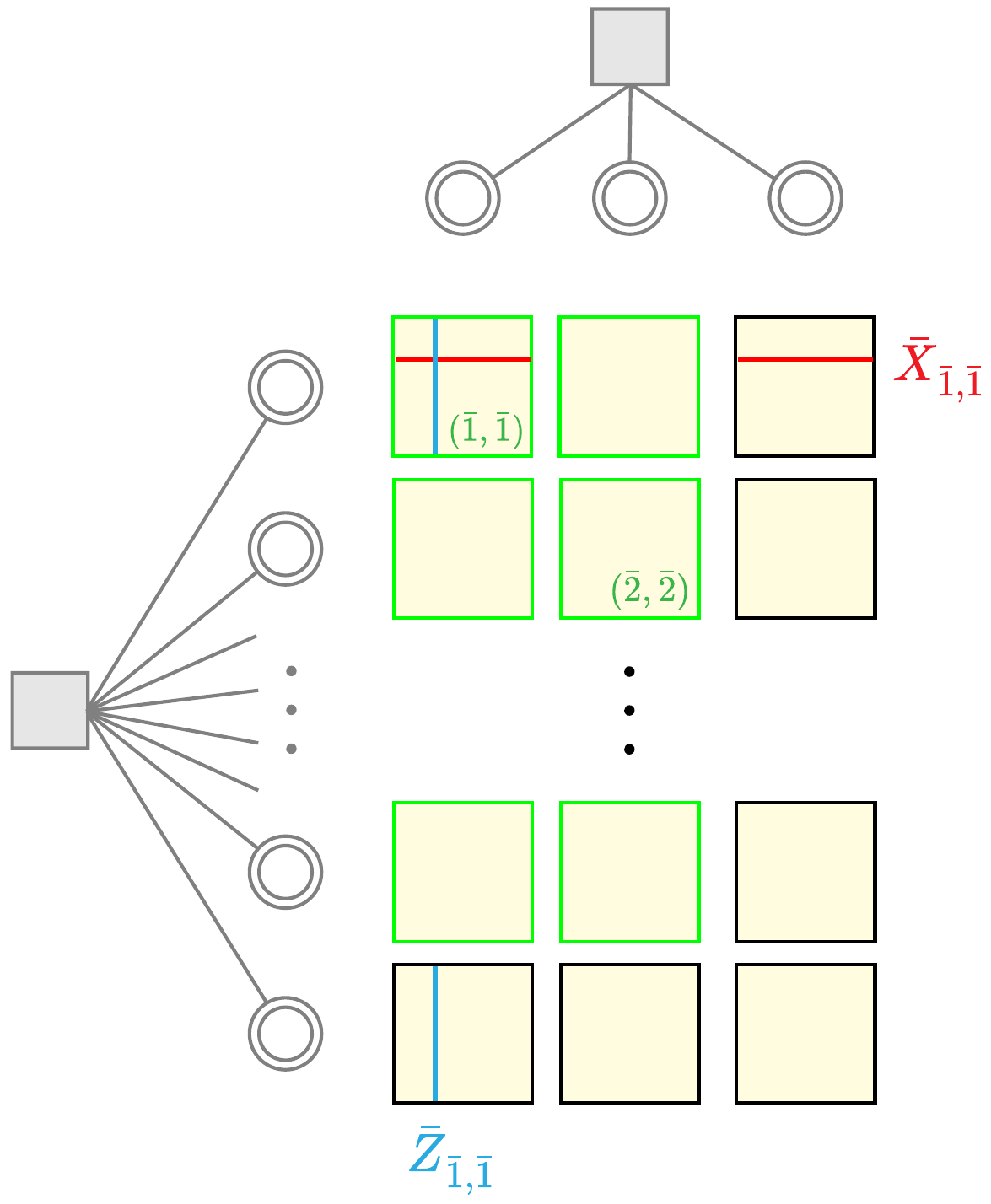}
    \caption{The qubit layout of the hypergraph product code with input concatenated 3-bit and $\ell$-bit parity codes is illustrated (before weight reduction for simplicity). Concentric circles denote the concatenated repetition-code bits, and yellow squares denote the corresponding surface-code patches. Highlighted green patches support logical $\bar{X}$ and $\bar{Z}$ intersections and hence index the logical qubits. Red and blue lines denote the support $\bar{X}$ and $\bar{Z}$ respectively for logical qubit $(\bar{1},\bar{1})$.}
    \label{fig:generalized GHZ layout}
\end{figure}

As detailed in \cite{LRESC}, the automorphisms of the above classical codes carry over when we concatenate our $[\ell,\ell-1,2]$ parity code with a repetition code. Specifically, any permutation of bits now become (segment-transversal) permutations of entire repetition-code segments. After the hypergraph product, these segment-transversal operations between repetition-code segments become patch-transversal operations between surface-code patches. In particular, we choose the $[(2\ell-3)c,\ell-1,2c]$ code to be the ``vertical'' code and the $[3c,2,2c]$ code to be the ``horizontal'' code in the hypergraph product according to Fig. \ref{fig:HGP detailed}. In analogy to Sec. \ref{sec:HGP and logicals}, we can choose our logical $\bar{X}$ and $\bar{Z}$ operators to intersect in a $(\ell-1) \times 2$ rectangle of surface-code patches and label our logical qubits according to these intersection patches using coordinates $(\bar{i},\bar{j})$, where $i=1,\dots,\ell-1$ and $j=1,2$; see Fig. \ref{fig:generalized GHZ layout} for an illustration. When we swap the last two rows of surface-code patches, we implement \eqref{eq:general parity swap} on the $(\cdot,\bar{1})$ and $(\cdot,\bar{2})$ logical $\bar{Z}$ operators simultaneously:
\begin{align}
    \bar{Z}_{\bar{i},\bar{j}} &\longrightarrow \bar{Z}_{\bar{i}} \bar{Z}_{\bar{\ell}-1,\bar{j}} \;,\;\; \forall i \neq \ell-1  \notag \\
    \bar{Z}_{\bar{\ell}-1,\bar{j}} &\longrightarrow \bar{Z}_{\bar{\ell}-1,\bar{j}} \, .
\end{align}
The logical $\bar{X}$ operators living on rows $\bar{1}$ through $\overline{\ell-2}$ remain unchanged since their corresponding data qubits were not involved in the row swap. The two logical $\bar{X}$ operators on row $\bar{\ell}-1$ transform into the product of all logical $\bar{X}$ operators in their corresponding columns:
\begin{align}
    \bar{X}_{\bar{i},\bar{j}} &\longrightarrow \bar{X}_{\bar{i},\bar{j}} \;,\;\; \forall i \neq \ell-1  \notag \\
    \bar{X}_{\bar{\ell}-1,\bar{j}} &\longrightarrow \prod_{q=1}^{\ell-1} \bar{X}_{\bar{q},\bar{j}} \, .
\end{align}
Indeed this is the only way these two logical $\bar{X}$s can transform in order to satisfy the Pauli algebra. Thus, we have shown that physically swapping the last two rows of surface-code patches implements two multi-target CNOT (fanout) gates on both columns of logical qubits with the $(\bar{\ell},\cdot)$ logical qubits as the control. As in the case of the $\llbracket 34,4,3 \rrbracket$ hypergraph product code in Sec. \ref{sec:HGP and logicals}, which uses the $[3,2,2]$ parity code, swapping appropriate columns of data qubits implements pairwise logical CNOT gates between logical qubits indexed by the same row.

Since the checks in the hypergraph product code remained invariant throughout the above manipulations, their combinations will remain invariant as well. As a result, the automorphisms survive through the quantum Tanner transformation described in Sec. \ref{sec:QTT}. After performing the QTT, the quantum code parameters become
\begin{align}
    n &= 3(2\ell-3)c^2  \notag \\
    k &= 2(\ell-1)  \notag \\
    d &= 2c \, .
\end{align}

\begin{figure}[t]
    \centering
    \includegraphics[width=0.7\linewidth]{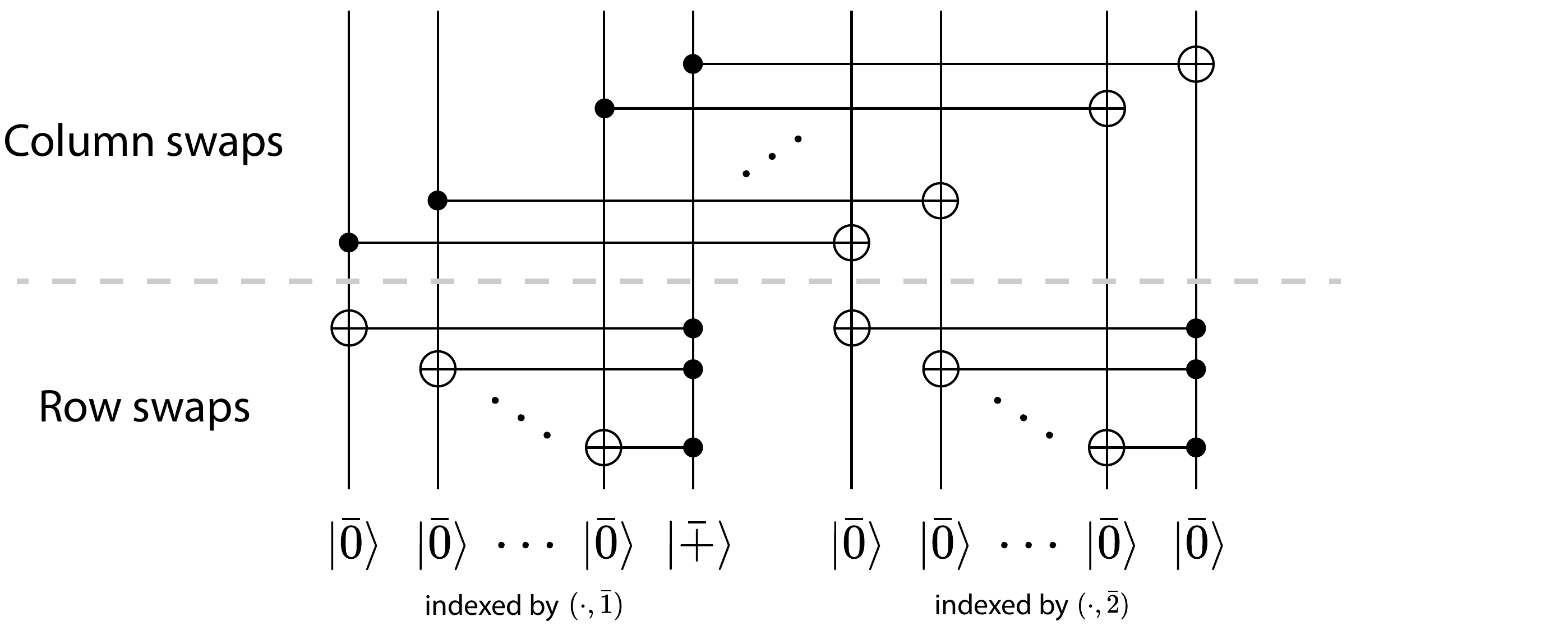}
    \caption{The logical circuit which prepares the logical GHZ state using the code described in Sec. \ref{sec:generalized GHZ}.}
    \label{fig:generalized logical GHZ circuit}
\end{figure}

We are now ready to describe the $2(\ell-1)$-qubit logical GHZ protocol using the quantum LDPC code described above.
\begin{enumerate}
    \item Transversally initialize the logical $\ket{\bar{0}}^{\otimes 2(\ell-1)}$ state.
    \item Apply $\bar{H}$ on logical qubit $(\bar{\ell},\bar{1})$. Or alternatively, measure its $\bar{X}$ operator and apply $\bar{Z}$ if the outcome is -1.
    \item Apply the aforementioned row and column patch-swaps to implement the logical circuit drawn in Fig. \ref{fig:generalized logical GHZ circuit}, which prepares the desired logical GHZ state.
\end{enumerate}

Lastly, we describe a modification to the concatenation step which allows for single-shot decoding at the expense of nonlocal connectivity. Instead of using a $[c,1,c]$ repetition code for each data bit in the parity code, we use a single logical bit in a $[c,\Theta(c),\Theta(c)]$ expander LDPC code \cite{Sipser_1996}. In order to achieve successful concatenation, we manipulate the expander code's parity-check matrix in systematic form $H' = (A \,|\, \ident)$ so that its generator matrix is given by $G = (\ident \,|\, A^\transpose)$. Coupling the first data bits of each expander code to the parity code then achieves the desired concatenation. The additional spatial overhead of using an expander code compared to a repetition code is given by the $O(1)$ constant prefactor in the $\Theta(c)$ code distance. Before when we used a repetition code for concatenation, the resulting hypergraph product code had the structure of surface-code patches connected by long-range checks at their boundaries. Now when we use an expander code for concatenation, the hypergraph product code will consist of quantum expander codes \cite{Leverrier_2015} linked by ``boundary'' checks. Importantly, quantum expander codes support single-shot decoding when the bit and check degrees in the input classical expander code are sufficiently large (but still $O(1)$) \cite{Fawzi_2018}. Thus, using expander LDPC codes instead of repetition codes for concatenation reduces the temporal overhead for error correction from $O(d)$ to $O(1)$ while only adding $O(1)$ spatial overhead, at the cost of nonlocal connectivity.

\bibliography{thebib}

\end{document}

%% file: preamble.tex
\usepackage{amsmath,amssymb,amsfonts,mathtools,dsfont,relsize}
\usepackage{graphicx}

\usepackage[colorlinks=true, urlcolor=blue, linkcolor=blue, citecolor=red, hyperindex=true, linktocpage=true]{hyperref}

\usepackage{amsthm}
\usepackage{qtree}

\usepackage{soul} % what is this?
\usepackage{bm}

\usepackage{enumerate}
\usepackage{stmaryrd} % what is this??

\renewcommand{\thesection}{\arabic{section}}
\renewcommand{\thesubsection}{\thesection.\arabic{subsection}}

\makeatletter
\renewcommand{\p@subsection}{}
\renewcommand{\p@subsubsection}{}
\makeatother

\newcommand{\vps}[0]{\vphantom{*}}

\DeclarePairedDelimiter{\expval}{\langle}{\rangle}

\DeclarePairedDelimiter{\ket}{\lvert}{\rangle}
\DeclarePairedDelimiter{\bra}{\langle}{\rvert}
\newcommand{\transpose}[0]{\mathsf{T}}

\newcommand{\ident}[0]{\mathds{1}}

\newcommand{\trace}{{\rm tr}}

\usepackage{qcircuit}

\makeatletter 
    
\renewcommand\onecolumngrid{% <<<<<<
\do@columngrid{one}{\@ne}%
\def\set@footnotewidth{\onecolumngrid}% <<<<<<<<<<<<<<<<
\def\footnoterule{\kern-6pt\hrule width 1.5in\kern6pt}%
}

\renewcommand\twocolumngrid{% <<<<<<
        \def\footnoterule{% restore rule
        \dimen@\skip\footins\divide\dimen@\thr@@
        \kern-\dimen@\hrule width.5in\kern\dimen@}
        \do@columngrid{mlt}{\tw@}
}%

\makeatother